\documentclass[final,5p,times,twocolumn,11pt]{elsarticle}

\usepackage{url}
\urldef{\ultraxurl}\url{http://www.ultrax-speech.org}

\usepackage{graphicx}
\graphicspath{{figures/}}

\usepackage{csquotes}
\usepackage{xcolor}
\usepackage{textcomp}
\usepackage{gensymb}
\usepackage{amsmath}
\usepackage{amssymb}
\usepackage{booktabs}
\usepackage{multicol}

\usepackage{hyperref}
\hypersetup{
    colorlinks = true,
    linkbordercolor = {white},
    linkcolor = {blue}
}

\journal{Speech Communication}
\biboptions{authoryear}

\begin{document}
\begin{frontmatter}

\title{Exploiting ultrasound tongue imaging for the\\automatic detection of speech articulation errors}

\author[a]{Manuel Sam Ribeiro}\corref{cor1}
\ead{sam.ribeiro@ed.ac.uk}

\author[b]{Joanne Cleland}
\ead{joanne.cleland@strath.ac.uk}

\author[a]{Aciel Eshky}
\ead{aeshky@ed.ac.uk}

\author[a]{Korin Richmond}
\ead{korin.richmond@ed.ac.uk}

\author[a]{Steve Renals}
\ead{s.renals@ed.ac.uk}

\address[a]{The Centre for Speech Technology Research, University of Edinburgh, UK}
\address[b]{Psychological Sciences and Health, University of Strathclyde, UK}

\cortext[cor1]{Corresponding author}

\begin{abstract}
Speech sound disorders are a common communication impairment in childhood.
Because speech disorders can negatively affect the lives and the development of children, clinical intervention is often recommended.
To help with diagnosis and treatment, clinicians use instrumented methods such as spectrograms or ultrasound tongue imaging to analyse speech articulations.
Analysis with these methods can be laborious for clinicians, therefore there is growing interest in its automation. 
In this paper, we investigate the contribution of ultrasound tongue imaging for the automatic detection of speech articulation errors.
Our systems are trained on typically developing child speech and augmented with a database of adult speech using audio and ultrasound.
Evaluation on typically developing speech indicates that pre-training on adult speech and jointly using ultrasound and audio gives 
the best results with an accuracy of 86.9\%.
To evaluate on disordered speech, we collect pronunciation scores from experienced speech and language therapists, focusing on
cases of velar fronting and gliding of /r/.
The scores show good inter-annotator agreement for velar fronting, but not for gliding errors.
For automatic velar fronting error detection, the best results are obtained when jointly using ultrasound and audio.
The best system correctly detects 86.6\% of the errors identified by experienced clinicians.
Out of all the segments identified as errors by the best system, 73.2\% match errors identified by clinicians.
Results on automatic gliding detection are harder to interpret due to poor inter-annotator agreement, but appear promising.
Overall findings suggest that automatic detection of speech articulation errors has potential to be integrated into ultrasound intervention software for automatically quantifying progress during speech therapy.
\end{abstract}


\end{frontmatter}

\section{Introduction}

Speech sound disorders (SSDs) are a common communication impairment in childhood \citep{wren2016prevalence}.
If left untreated, SSDs can have a negative impact on the social and emotional development of children and can lead to poor educational outcomes.
For example, self-awareness of disordered speech contributes to low confidence in social situations when children engage with their peers or educators. In turn, this introduces communication barriers that lead to lower literacy levels  \citep{johnson2010twenty, lewis2011literacy, mccormack2011nationally}.

It is estimated that SSDs affect between 2.3\% and 24.6\% of children \citep{law2000prevalence, wren2016prevalence}.
Speech and language therapy is often recommended, with the majority of interventions heavily reliant on auditory feedback. That is, the speech and language therapist (SLT) relies on their perceptual skills to give the child verbal feedback during intervention; and in turn the child relies on their perceptual skills to modify their articulations. This may also be accompanied by auditory cues describing where and how to place the articulators to produce the target sound.
Interventions are often successful, especially for younger children \citep{mcleod2020waiting}.
However, some children do not respond well and the SSD becomes persistent.
There is growing evidence that including visual biofeedback (VBF) during therapy is beneficial for such children \citep{sugden2019systematic}.
VBF allows the visualization of the vocal tract during the speech production process, enabling children to view articulations in real-time.

With the widespread use of technology, there is increasing interest in automatically processing speech therapy tasks.
This type of automation can be helpful to teachers and parents, who may use screening tools to determine the presence or absence of SSDs \citep{sadeghian2015towards, ward2016automated}; and to clinicians, who can save time on these time-consuming tasks \citep{ribeiro2019ultrasound}.
Clinicians and researchers are trained to use instrumented methods such as spectrograms or ultrasound tongue imaging to assist in the assessment, diagnosis, or quantification of treatment efficacy.
These methods, however, can be laborious and impractical in the speech therapy clinic, as they still rely on manual annotation by the therapist or other trained professionals.
Typical tasks include the identification of utterances spoken by the child, the identification of boundaries of target words, or measurements to determine correctness of speech articulations.

In this paper, we are concerned with the \emph{automatic detection of speech articulation errors for speech therapy using ultrasound visual biofeedback}.
This is intended as a tool for clinicians to automatically process data from ultrasound visual biofeedback assessment and therapy sessions.
This work provides the following contributions:
1) the collection and analysis of pronunciation scores of velar fronting and gliding of /r/ given by experienced speech and language therapists;
2) a method for the automatic detection of speech articulation errors to be used by clinicians when processing data collected after therapy sessions;
3) an investigation of the impact of ultrasound tongue imaging for automatic error detection; and
4) an analysis of the impact of out-of-domain adult speech data for automatic error detection.
Our method is evaluated on typically developing child speech and, more specifically, on cases of velar fronting and gliding of /r/ in Scottish English child speakers.
These two errors are common in children with SSDs and amenable to intervention with ultrasound visual biofeedback \citep{sugden2019systematic}.

Section \ref{sec:background} provides background on speech disorders and recent evidence on the benefits of ultrasound visual biofeedback, as well as a review of recent literature on automatic speech articulation error detection.
The data used throughout this work is described in Section \ref{sec:data}.
Section \ref{sec:expert_detection} describes a perceptual experiment where SLTs were asked to rate the goodness of pronunciation of phone instances, using ultrasound and audio data.
Section \ref{sec:automatic_detection} describes a set of experiments for the automatic scoring of phone pronunciations using ultrasound tongue imaging.
Finally, Sections \ref{sec:discussion} and \ref{sec:conclusion} provide an overall discussion and conclusion for this work, respectively.

\section{Background}
\label{sec:background}

\subsection{Speech sound disorders}
SSDs occur when children exhibit difficulties in the production of speech sounds in their native language.
\emph{Organic speech sound disorders} denote difficulties that are associated with known causes.
These causes may be motor or neurological (e.g. childhood dysarthria associated with cerebral palsy), structural (e.g. cleft lip and palate), or sensory (e.g. hearing impairments).
\emph{Functional speech sound disorders} are related to difficulties producing intelligible or acceptable speech with unknown causes.
These disorders may be associated with motor production (articulation or motor speech disorders), or related to predictable or rule-based errors (phonological disorders) \citep{asha2020}.

\begin{table}[t]
\centering
\resizebox{\columnwidth}{!}{%
\begin{tabular}{@{}lll@{}}
\toprule
\textbf{Substitution} & \textbf{Description}                         & \textbf{Example} \\ \midrule
Fronting              & Alveolars (/t,d/) replace velars (/k,g/)     & \emph{cookie} \textrightarrow{} \emph{tootie}   \\
Backing               & Velars (/k,g/) replace alveolars (/t,d/)     & \emph{dog} \textrightarrow{} \emph{gog}          \\
Gliding               & Glides (/w,j/ replace liquids (/r,l/)        & \emph{rabbit} \textrightarrow{} \emph{wabbit}    \\
Stopping              & Stops (/p, d/) replace fricatives (/f,s/)    & \emph{zoo}\textrightarrow{} \emph{doo}         \\
Labialisation         & Labials (/p,b/) replace non-labials          & \emph{tie} \textrightarrow{} \emph{pie}         \\ 
\bottomrule
\end{tabular}%
}
\caption{Selected examples of substitutions. A phone or group of phones is systematically replaced by another phone or group of phones.}
\label{tab:background-substitutions}
\end{table}

Both types of SSDs can result in a variety of speech patterns.
\textbf{Substitutions} are a common pattern where a phone or a group of phones is replaced by another phone or group of phones.
Table \ref{tab:background-substitutions} summarises common substitutions.
We highlight here the two processes that are relevant to this work.
\emph{Fronting} occurs when phones that are produced towards back of the mouth (velars such as /k, g/) are replaced with phones produced towards the front of the mouth (alveolars such as /t, d/).
This leads to word instances such as \emph{cookie} \textrightarrow{} \emph{tootie} or \emph{gate} \textrightarrow{} \emph{date}.
\emph{Gliding} occurs when liquids (e.g. /r, l/) are replaced with glides (e.g. /w/), originating cases such as \emph{rabbit} \textrightarrow{} \emph{wabbit} or \emph{leg} \textrightarrow{}\emph{weg}. 
Other examples of substitutions not listed in Table \ref{tab:background-substitutions} are affrication, vowelization, depalatalization, or alveolarization \citep{mcleod2017children}.

Beyond substitutions, we may observe \textbf{insertions} and \textbf{deletions} of phones in words (e.g. \emph{black} \textrightarrow{} \emph{buhlack}, \emph{spoon} \textrightarrow{} \emph{poon}).
Alternatively, \textbf{assimilation} denotes cases when specific sounds are transformed due to the influence of those around it.
For example, the process of nasal assimilation occurs when a non-nasal sound becomes nasal due to the presence of a nasal sound in the word (e.g. \emph{bunny}\textrightarrow{} \emph{nunny}).
Similarly, pre-vocalic voicing occurs when a voiceless phone becomes voiced when followed by a vowel (e.g. \emph{comb} \textrightarrow{} \emph{gomb}).
Additional phonological patterns may be influenced by \textbf{syllable structure}.
For example, cluster reduction, where a consonant cluster is reduced (\emph{plane}\textrightarrow{} \emph{pane}, \emph{clean} \textrightarrow{} \emph{keen}); the deletion of a consonant at the beginning (\emph{bunny} \textrightarrow{} \emph{unny}) or end of the syllable (\emph{bus} \textrightarrow{} \emph{bu}); or the deletion of weak or unstressed syllables in words (\emph{banana} \textrightarrow{} \emph{nana}).
Other \textbf{phonetic distortions} may also be observed (for example, a lateral /s/).

Many of these processes are typical stages in the speech development of children.
They are, however, expected to be eliminated as children reach a certain age.
For instance, the typical age of elimination of velar fronting is around three years \citep{mcleod2018children}. 
On the other hand, because /r/ in particular is late acquired, gliding of this phone is typically eliminated around the age of five \citep{mcleod2018children}. 
Children that persist with one or more of these processes beyond their expected age of elimination usually require speech and language therapy.

\subsection{Ultrasound visual biofeedback}

\begin{figure}
\includegraphics[width=\columnwidth]{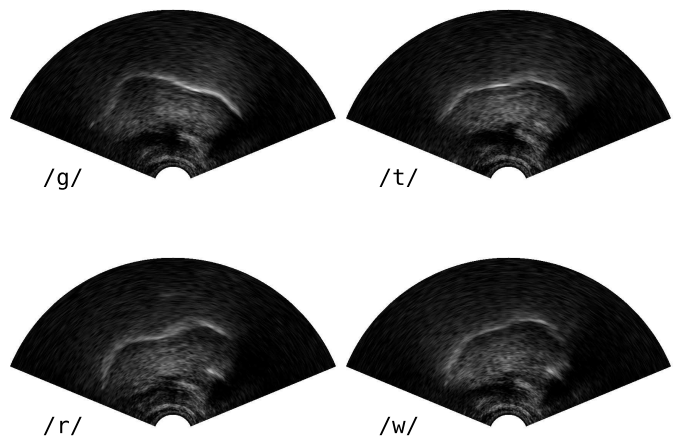}
\caption{Ultrasound tongue images collected from a typically developing speaker (female, aged 11). Each frame is the mid-point of the phone showing a midsaggital view of the oral cavity with the tip of the tongue facing right. Samples extracted from the Ultrasuite repository \citep{eshky2018ultrasuite}.}
\label{fig:td_samples}
\end{figure}

In the context of speech sound disorders, visual biofeedback involves the use of instrumented methods to provide visual information regarding the position, movement, or shape of intra-oral articulators during speech production \citep{sugden2019systematic}.
Common techniques to provide real-time visual biofeedback for speech therapy are electropalatography (EPG, \cite{lee2009electropalatography}), electromagnetic articulography (EMA, \cite{katz2010treating}), and ultrasound tongue imaging (UTI, \cite{sugden2019systematic}).
EPG uses an artificial palate to measure the contact points between the tongue and hard palate.
However, the manufacture of custom-made palates incurs additional costs and may limit the use of this technique to a few patients.
EMA requires the placement of sensor coils on the tongue and other articulators to measure their position over time, which can be both expensive and intrusive for children.
Ultrasound tongue imaging uses diagnostic ultrasound operating in B-mode to visualise the tongue surface during the speech production process.
A real-time B-mode ultrasound transducer is placed under the speaker's chin to generate a mid-saggital or coronal view of the tongue.
This form of ultrasound is  clinically safe, non-invasive, non-intrusive, portable, and relatively cheap \citep{stone2005guide}.
Figure \ref{fig:td_samples} provides examples of ultrasound images of the tongue for a typically developing speaker.

Increasing evidence shows that ultrasound VBF can be beneficial for patients, therapists, and annotators \citep{bernhardt2005ultrasound, cleland2019enabling, cleland2020impact}.
U-VBF is beneficial when used in intervention for a range of speech sound disorders, particularly if used in the initial stages of motor learning \citep{sugden2019systematic}.
Related work suggests that U-VBF can be used as an objective measure of progress in intervention \citep{cleland2020dorsal},
or to complement audio feedback and contribute to positive reinforcement \citep{roxburgh2015articulation}.
U-VBF can also assist annotators in the identification of covert errors and increase inter-annotator agreement scores \citep{cleland2020impact}.
Additionally, U-VBF can contribute to the automatic processing of speech therapy recordings.
Recent work used ultrasound data to develop tongue contour extractors \citep{fabre2015tongue}, animate a tongue model \citep{fabre2017automatic}, automatically synchronise therapy recordings \citep{eshky2019synchronising}, and for speaker diarisation and alignment of therapy sessions \citep{ribeiro2019ultrasound}.
There are, however, several challenges associated with the automatic processing of ultrasound tongue images \citep{stone2005guide, ribeiro2019speaker}.
Ultrasound output tends to be noisy, with unrelated high-contrast edges, speckle noise, or interruptions of the tongue surface.
Image quality may also be affected by speaker characteristics (e.g. age and physiology) or session variability (e.g. incorrect or variable probe placement).

\subsection{Automatic speech error detection}

Automatic speech error detection aims to identify inaccurate productions of words or phones.
These are often described in terms of insertions, deletions, and substitutions.
Most studies adopt techniques from computer assisted pronunciation training, primarily developed for adult speakers using language learning systems (e.g. \cite{witt2000phone, witt2012automatic, hu2015improved}).
This work is often considered part of the broader area of Computer Assisted Language Learning (CALL, \cite{beatty2013teaching}).
In children, speech error, or mispronunciation, detection can be used to assess reading levels (e.g. \cite{black2010automatic}, \cite{proencca2018mispronunciation}), or with disordered speech for Computer Assisted Speech Therapy (CAST, e.g. \cite{saz2009tools, parnandi2015development, ahmed2018speech}).
CALL systems are generally concerned with a global pronunciation score, which may or may not use the speaker's native language, while CAST systems aim to identify error types or underlying phonological processes.

Mispronunciation detection systems often use speech recognition techniques to compute pronunciation scores.
Training data for the acoustic models is L2 speech for language learning, or typically developing speech for therapy applications.
Other sources, if available, may consist of in-domain speech from the learners' native language or disordered speech.
In-domain data can be used to develop extended search lattices accepting non-canonical pronunciation alternatives \citep{harrison2009implementation, ward2016automated, dudy2018automatic}.
The trained acoustic models and extended transducers then decode unseen utterances for which the text is known.
To provide pronunciation scores, \emph{likelihood-based systems} use the log-likelihoods generated by the models.
The Goodness of Pronunciation score (GOP, \cite{witt2000phone}) is a widely-used method for such systems.
In its simplest form, the GOP score is the log-likelihood ratio between a target phone and a competing phone.
Recently, Gaussian mixture models have been replaced by deep neural networks, with GOP-like scores defined over neural network posteriors \citep{hu2015improved}.
Alternatively, \emph{classifier-based systems} use model outputs with supervised classifiers to determine error types or to provide more informed feedback.
However, these methods require supervised in-domain data (e.g. phone-level annotated disordered speech), which are costly to acquire.

When annotated in-domain data is not available, a possible approach is to learn distributions over canonical training examples, such as  typically developing speech.
Unseen samples can then be compared against those distributions.
\cite{shahin2018anomaly} use one-class support vector machines to model the distribution of features describing manner and place of articulation.
\cite{wang2019child} use Siamese recurrent networks, with positive and negative samples drawn from typically developing speech.
In this paper, we adopt a similar strategy.
Because we do not have annotated disordered speech for training, the acoustic model is trained only on typically developing speech.
Scoring is based on a GOP-like score defined over neural network posteriors.
Section \ref{sec:automatic_detection} provides additional details on our model implementation.

\section{Data}
\label{sec:data}

We use data from the \textbf{Ultrasuite repository}\footnote{\label{fn:ultrasuite}\url{https://www.ultrax-speech.org/ultrasuite}} \citep{eshky2018ultrasuite}, consisting of synchronised ultrasound and audio data from child speech therapy sessions.
Ultrasuite currently contains three datasets of child speech.
Ultrax Typically Developing (UXTD) includes recordings of 58 typically developing children.
The remaining datasets include recordings from children with speech sound disorders collected over the course of assessment and therapy sessions: Ultrax Speech Sound Disorders  (UXSSD,  8  children)  and  Ultraphonix  (UPX,  20  children). 
Assessment sessions denote recordings at various stages of therapy: baseline (before therapy),  mid-therapy,  post-therapy  (immediately after therapy), and maintenance (several months after therapy).
For the child speech datasets, ultrasound was recorded with an Ultrasonix SonixRP machine using Articulate Assistant Advanced (AAA, \cite{articulate2010articulate}) software at $\sim$120fps with a 135\degree ~field of view.
A single B-Mode ultrasound frame has 412 echo returns for each of 63 scan lines, giving a $63\times412$ \enquote{raw} ultrasound frame capturing a mid-sagittal view of the tongue.
Samples from this data are illustrated in Figure \ref{fig:td_samples}.

To complement the Ultrasuite repository, we use the \textbf{Tongue and Lips corpus}\footnote{Available via the Ultrasuite Repository, see footnote \ref{fn:ultrasuite}.}(TaL, \cite{ribeiro2021tal}).
TaL is a corpus of synchronised ultrasound, audio, and lip videos from 82 adult native speakers of English.
Ultrasound in the TaL corpus was recorded using Articulate Instruments' Micro system \citep{articulate2010articulate} at $\sim$80fps with a 92\degree ~field of view.
TaL used a different transducer than the one in the Ultrasuite data collection.
Because of this, an ultrasound frame of the TaL corpus contains  842  echo  returns  for  each  of  64  scan  lines ($64\times842$ \enquote{raw} ultrasound frame).

\section{Expert speech error detection}
\label{sec:expert_detection}

In this section, our \textbf{goal} is the collection of pronunciation scores for speech segments produced by children with speech sound disorders.
This data is to be used in the evaluation of the automatic error detection systems described in Section \ref{sec:automatic_detection}.
We recruited Speech and Language Therapists with experience using ultrasound visual biofeedback 
and who routinely work with Scottish English-speaking children.
The collection of these scores aimed to simulate the process therapists undergo after collecting data from speech therapy sessions.
We are interested in the processes of velar fronting and gliding of /r/, therefore we focused on productions of /k, g/ (\textit{velars}) and pre-vocalic /r/ (\textit{rhotic}).
We expected SLTs to be able to identify correct and incorrect productions of velars and rhotics with good reliability.

\subsection{Data preparation}

We used data from the eight children available in Ultrasuite's UXSSD dataset.
Children in this dataset were treated for velar fronting, therefore we expected to observe an increasing number of correct velar productions throughout assessment sessions.
Children's response to intervention is reported in \citet{cleland2015using}.
Because intervention for these children did not focus on correcting rhotic productions, this led to an imbalanced set of correct and incorrect rhotic samples.
We pre-selected words containing the target velar and rhotic phones and occurring within prompts of type \enquote{A} (single words) in assessment sessions (baseline, mid-therapy, post-therapy, and maintenance).
We discarded words that contained more than one instance of a target phone (e.g.\ \enquote{cake}) or corrupted word instances (e.g.\ overlapping or unintelligible speech, other background noise, etc).
From the pre-selected word list, we randomly sampled 96 word instances per child.
Samples were balanced across assessment sessions and across velars and rhotics.
For each child, an assessment session contained 24 word instances (12 velars and 12 rhotics).
Where possible, samples were also balanced for the position of the target phone in the word (initial, medial, or final).
The final set of samples consisted of a total of 768 word instances with a vocabulary of 148 words, which
we denote as the \textbf{main set}.

We generated a set of additional samples from one of the speakers available in Ultrasuite's UPX dataset.
We selected speaker 04M, treated for velar fronting and reported to have good improvement after intervention \citep{cleland2019enabling}.
As the speakers in the main set, this speaker was also not treated for gliding.
The sampling process was repeated and a total of 24 word instances were selected (12 velars and 12 rhotics), balanced across assessment sessions.
We denote this set of samples the \textbf{control set}.

\begin{figure}
\includegraphics[width=\columnwidth]{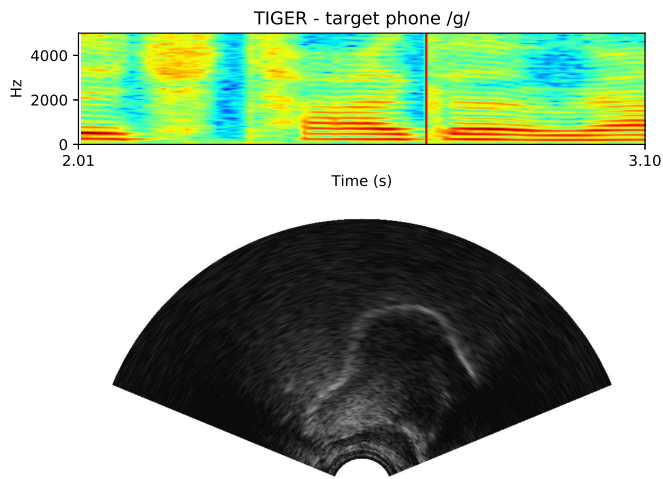}
\caption{\label{fig:annotation-video-sample} Video frame for the word \enquote{tiger} produced by speaker 02M in a post-therapy assessment session. The annotator is requested to score the target phone /g/ in a word medial position. The vertical red line in the spectrogram indicates the current position of the video.}
\end{figure}

\subsection{Method}
\label{subsec_expert_method}

To annotate the word instances, we recruited 8 annotators.
All annotators were SLTs with more than 4 years of experience and who routinely work with children speaking Scottish English.
Additionally, the annotators had at least 3 years of experience with ultrasound visual biofeedback, with two annotators having more than 10 years of experience.
Each annotator was assigned 96 words from the main set and the 24 words from the control set.
Words taken from the main set were selected such that they were produced by a single child.
Therefore, each SLT annotated data from two children (one main and one control).
For intra-annotator agreement, 20\% of the words (24 samples) were repeated in the annotation list.
Of these, 12 were taken from the control set and 12 from the main set.
Each SLT annotated a total of 144 word instances.

Results were collected via a web interface displaying a video of each word sample separately.
The video contained the spectrogram, ultrasound images of the tongue, and the audio for each word.
Figure \ref{fig:annotation-video-sample} illustrates one video frame extracted from one of the samples.
Annotators were allowed to play videos at normal, half, or quarter speed up to a maximum of 6 total playbacks. 

\begin{figure}
\includegraphics[width=\columnwidth]{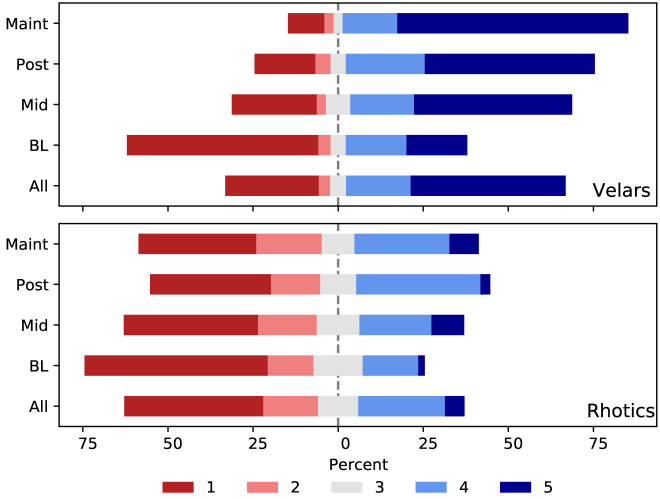}
\caption{\label{fig:annotation-bar-chart} Normalised distribution of primary scores for Velars and Rhotics, ordered chronologically (bottom up) by assessment session: baseline (BL), mid-therapy (Mid), post-therapy (Post), and maintenance (Maint).}
\end{figure}

Annotators were instructed to rate the target phone on a 5-point Likert scale,
where 1 indicates \emph{wrong pronunciation} and 5 indicates \emph{perfect pronunciation}.
The first question requested a score for the target phone (e.g. \enquote{Please rate the velar /k/ in the sample}).
This score is denoted the \textbf{primary score}.
If the annotator scored 3 or lower in the primary score, we requested a \textbf{secondary score}.
The secondary score asked the annotator to rate the target phone with respect to an expected substitution.
(e.g. \enquote{Please rate the target phone for alveolar substitution} or \enquote{Please rate the target phone for gliding substitution}).
An optional field allowed annotators to provide a short comment for each sample.

Given the primary score $s_p$ and the secondary score $s_s$, we determine a \textbf{combined score} $s_c$ defined as $s_c = \log(s_p)-\log(s_s)$.
Because we did not request a secondary score when perfect pronunciation was rated for the target phone, we assumed a value of 1 for $s_s$ when computing the combined score.
The score $s_c$ is positive if there is a preference for the primary class (e.g. velars or rhotics) and negative if there is a preference for the secondary class (e.g. alveolars or glide).
If the annotator gave the same primary and secondary scores, then this uncertainty is represented in $s_c$ with 0.
This preference can be further simplified to produce a \textbf{binary score} $s_b$.
For each sample, positive values of the combined scores are treated as correct productions of the primary class and negative or zero values as incorrect productions.

\subsection{Results}
\label{subsec:annotation_results}

Figure \ref{fig:annotation-bar-chart} shows the normalised distribution of the primary score for all annotated samples, ordered chronologically by session.
The number of incorrect velars across assessment sessions decreases over time, while the distribution for rhotics is more or less stable.
This is expected since intervention for these children focused on velar fronting and production of /r/ was not addressed.

\begin{figure}[t]
\includegraphics[width=\columnwidth]{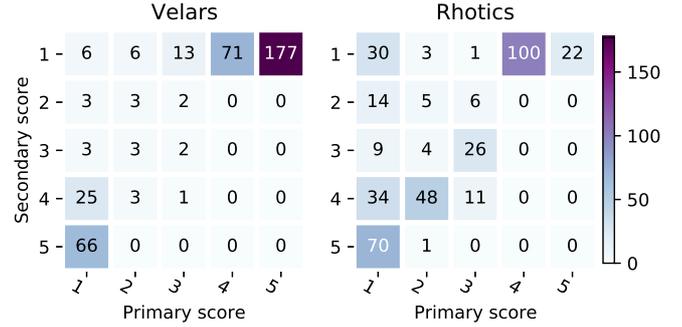}
\caption{\label{fig:annotation-primary-secondary-matrix} Frequency of primary and secondary scores given by annotators for Velars and Rhotics in the main set with duplicate entries removed, where 1 indicates wrong pronunciation and 5 indicates perfect pronunciation.}
\end{figure}

Figure \ref{fig:annotation-primary-secondary-matrix} shows the frequency of primary and secondary scores for velars and rhotics in the main set. 
We remove duplicate samples used for intra-annotator agreement, keeping the score of the first sample to be rated.
For this work, we are primarily interested in the correct production of velars and rhotics and clear cases of substitutions (fronting and gliding).
Cases of correct pronunciations for the expected class are identified by a high primary score (4 or 5).
Cases of velar fronting or gliding are identified by a low primary score (1 or 2) and a high secondary score (4 or 5).
We observe from Figure \ref{fig:annotation-primary-secondary-matrix} that 342 out of the 384 velar samples (89.06\%) fall under one of these two cases.
Of these samples, 248 are correct velars and 94 are alveolar substitutions.
Rhotics include a smaller number of correct productions or gliding (275 out of 384 samples, 71.61\%).
Of these, 122 are marked as a correct production of /r/, while 153 denote cases of gliding.

\begin{figure*}
\centering
\includegraphics[width=0.95\textwidth]{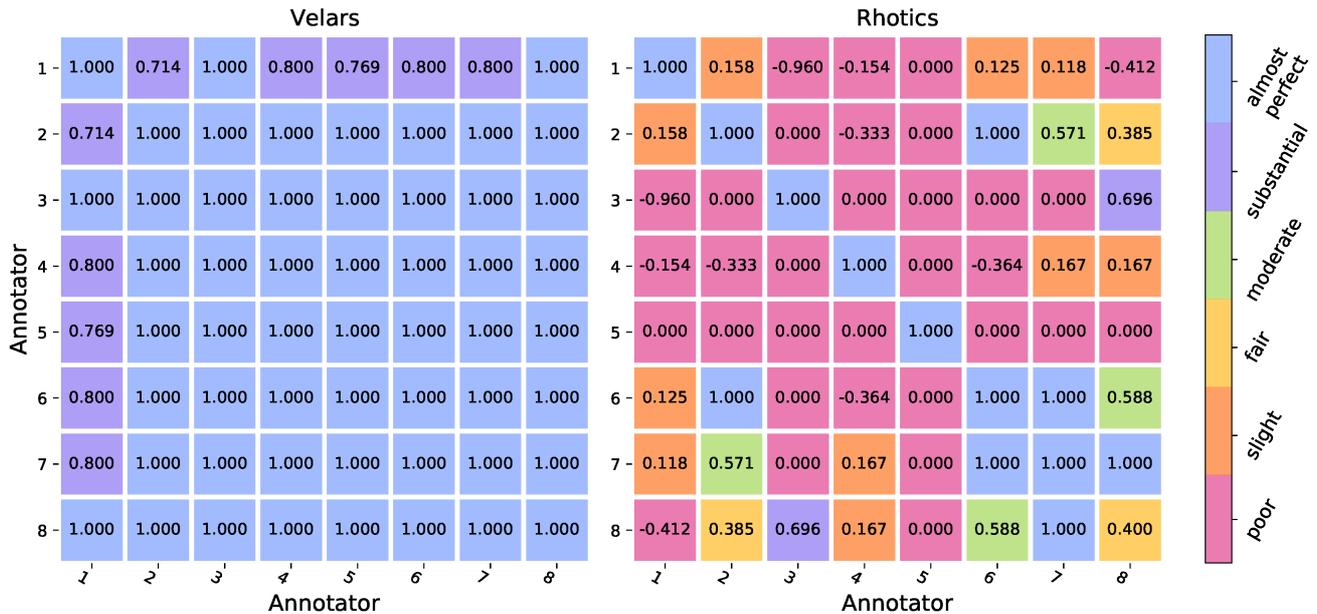}
\caption{\label{fig:annotation-cohen-kappa} Pairwise Cohen's $\kappa$ for the binary score, excluding cases not rated as correct productions or clear substitutions. Diagonal values show $\kappa$ for intra-annotator agreement, while off-diagonal shows pairwise inter-annotator agreement. Each cell is colour-coded according to the levels of agreement proposed by \citet[pp 164-165]{landis1977measurement}.}
\end{figure*}

Some annotators used the optional comment field to elaborate on their score, particularly for incorrect cases that were not instances of velar fronting or gliding.
For velars, some of the cases were reported to be uvular, palatal, or postalveolar realisations, or omitted phones.
For rhotics, most of the non-typical scores indicated deletion of /r/ with some cases reporting a distortion towards a labiodental approximant.

\subsection{Annotator agreement}
\label{subsec:annotator-agreement}

We compute inter-annotator agreement using the control set of samples, rated by all annotators.
Duplicates are removed by choosing the first rating and discarding the second.
Intra-annotator agreement is computed on the 20\% duplicate samples, half from the main set and half from the control set.

To measure global agreement, we use \emph{Krippendorf's $\alpha$} \citep{krippendorff2004content}, which computes annotator 
agreement for multiple annotators and supports several levels of measurements.
We compute $\alpha$ using a difference function for ordinal data for the primary score, a function for interval data for the combined score, and a function for nominal data for the binary score \citep{krippendorff2011computing}.
According to \citet[pp 241-243]{krippendorff2004content}, $\alpha > 0.8$ indicates reliable data, while $0.667 \leqslant \alpha \leqslant 0.8$ indicates moderately reliable data. When $\alpha < 0.667$, the data should be considered unreliable.
Table \ref{tab:annotation-krippendorf} shows $\alpha$ values for primary, combined, and binary scores.
For the binary score, we exclude samples not rated as correct productions or clear substitutions.
Overall annotator agreement appears to be very good for velar samples and poor for rhotic samples.

\begin{table}
\centering
\resizebox{0.65\columnwidth}{!}{%
\begin{tabular}{@{}cccc@{}}
\toprule
            & \textbf{Primary} & \textbf{Combined} & \textbf{Binary} \\ \midrule
All         &  0.601   &  0.578 & 0.579 \\
Velars      &  0.883   &  0.868 & 0.946 \\ 
Rhotics     &  0.210   &  0.117 & 0.050 \\
\bottomrule
\end{tabular}%
}
\caption{Krippendorf's $\alpha$ for primary score $s_p$, combined score $s_c$, and binary score $s_b$ computed for all annotators across all, velar, and rhotic samples. Agreement for binary score excludes samples not rated as correct productions or clear substitutions.}
\label{tab:annotation-krippendorf}
\end{table}

We measure pairwise agreement using Cohen's $\kappa$ \citep{cohen1960coefficient}, which measures agreement between two raters on categorical data.
The $kappa$ statistic is a standardised metric where $\kappa \in [-1, 1]$, with $\kappa = 0$ denoting chance agreement and $\kappa=1$ denoting perfect agreement.
Traditionally, Cohen's $\kappa$ is discussed according to the five agreement levels suggested by \citet[pp 164-165]{landis1977measurement}.
These group values of $\kappa$ into: poor ($\kappa \leq 0.0$), slight ($0.0 < \kappa \leqslant 0.2$), fair ($0.2 < \kappa \leqslant 0.40$), moderate ($0.40 < \kappa \leqslant 0.60$), substantial ($0.60 < \kappa \leqslant 0.80$), and almost perfect ($0.80 < \kappa \leqslant 1.0$) agreement.
Figure \ref{fig:annotation-cohen-kappa} visualises pairwise annotator agreement for the binary score.
Off-diagonal values denote pairwise inter-annotator agreement on the control set, whereas diagonal values denote intra-annotator agreement on the duplicate samples.

According to Figure \ref{fig:annotation-cohen-kappa}, scores provided for the velar samples are very consistent and reliable, with perfect agreement across most raters.
This is observed for inter and intra-annotator scores.
Annotator 1 has substantial agreement with some of the other raters, but not perfect.
Excluding samples rated by annotator 1 leads to improved global inter-annotator agreement for velar samples on the combined score ($\alpha=0.922$).
Results for the rhotic samples, however, indicate a substantial disagreement between annotators.
We observe a perfect agreement for intra-annotator scores across all annotators except annotator 8.
Rhotic agreement between annotators 6, 7, and 8 is higher than other raters, with perfect or moderate agreement.
However, considering only those three raters, global agreement on the combined score is still lower than the moderate reliability threshold for Krippendorf's $\alpha$ ($\alpha=0.631$).
There are various reasons that could explain the agreement discrepancy between velar and rhotic samples.
We provide further insights into these results in Section \ref{sec:discussion}.
However, these results indicate that we should use rhotic scores carefully when evaluating automatic error detection systems in the next section.

\section{Automatic speech error detection}
\label{sec:automatic_detection}

In this section we investigate the automatic detection of speech errors in typically developing and disordered Scottish English child speech.
The proposed system is designed as a tool to be used by Speech and Language Therapists on data collected from speech therapy and assessment sessions.
Therefore, we evaluate model scores with the expert scores for velar fronting and gliding of /r/ provided by therapists in Section \ref{sec:expert_detection}.
Our \textbf{goal} is to investigate the proposed system's ability to simulate expert behaviour in the detection of substitution errors.
Additionally, we aim to analyse the contribution of \emph{ultrasound tongue imaging} and \emph{out-of-domain adult data} on the automatic speech error detection.

\subsection{Data preparation}

\begin{table}[]
\centering
\resizebox{1.0\columnwidth}{!}{%
\begin{tabular}{@{}ccccl@{}}
\toprule
\multicolumn{1}{c}{\textbf{Data set}} & \multicolumn{1}{c}{\textbf{Source}} & \multicolumn{1}{c}{\textbf{Speakers}} & \multicolumn{1}{c}{\textbf{Samples}} & \multicolumn{1}{c}{\textbf{Notes}}                                                                        \\ \midrule
Train                                 & UXTD                                & 45                                    & 8302                                 & Child in-domain train data                                                                                \\
Train                                 & TaL                                 & 81                                    & 81193                                & Adult out-of-domain train data                                                                                  \\
Validation                            & UXTD                                & 5                                     & 534                                  &
Child in-domain validation data                                                                              \\
Test                                  & UXTD                                & 13                                    & 901                                  & Typically developing evaluation set                                                                       \\
Test                                  & UXSSD                               & 8                                     & 768                                  & Disordered speech evaluation set \\ \bottomrule
\end{tabular}%
}
\caption{Datasets used for automatic speech error detection. All sets contain samples from all places of articulation, except the disordered speech evaluation set, which contains the velar (384) and rhotic (384) samples rated by annotators in Section \ref{sec:expert_detection}.}
\label{tab:model-data-sets}
\end{table}

For \textbf{training data}, we use Ultrax Typically Developing (UXTD), which collected data from 58 child speakers.
UXTD includes a subset of utterances with manually-annotated word boundaries for 13 speakers.
We save data from those speakers for evaluation.
From the remaining 45 speakers, we randomly select 40 speakers for training and 5 speakers for validation.
The TaL corpus of adult speech is used as an additional source of training data.
We use the TaL80 dataset, containing data from 81 speakers.

For \textbf{evaluation data}, we use an evaluation set of disordered speech samples and an evaluation set of typically developing speech samples.
The disordered speech samples consist of the \emph{main set} rated by expert SLTs, described in Section \ref{sec:expert_detection}.
This is a set of 768 word instances from the UXSSD dataset.
The typically developing samples are extracted from the UXTD dataset, which includes 220 utterances with manually annotated word boundaries.
These utterances are produced by 13 speakers, disjoint from those in the training and validation sets.
After pre-processing, the typically developing evaluation set consists of 901 phone instances extracted from 866 words with a vocabulary of 153 words.

Because output classes are unbalanced, we control the number of samples per class for the training data.
For classes that are under-represented, we retrieve additional examples.
This is done by perturbing the anchor frame by up to 40ms for under-represented classes.
For classes that are over-represented, we randomly sample 1000 and 10000 examples for the UXTD and TaL sets, respectively.
After balancing and pre-processing, we have a total of 8302 for the UXTD training data.
The TaL corpus is larger than UXTD and has a total of 81193 samples.
Table \ref{tab:model-data-sets} shows the datasets used in this section, and their respective number of speakers and number of samples.

\begin{figure}
\includegraphics[width=\columnwidth]{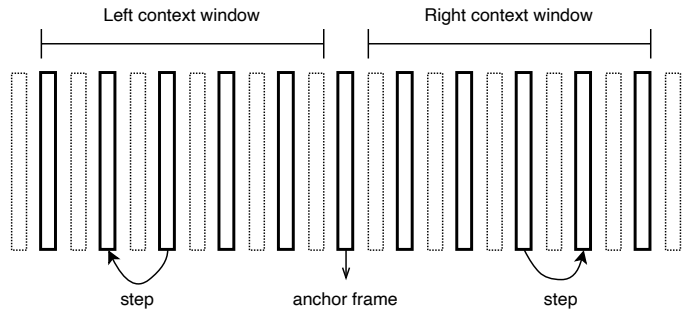}
\caption{\label{fig:diagram-sample-build} Sample build process. A sample is constructed around an anchor frame, typically the mid-point of a phone instance. Context windows are fixed at 100ms. Step is set such that 5 context frames are extracted for audio and 4 context frames for ultrasound.}
\end{figure}

The Kaldi speech recognition toolkit \citep{povey2011kaldi} is used to force-align all datasets at the phone level using the reference audio.
For the evaluation data, we constrain the phone alignment to the manually verified word boundaries.
The phone set is a Scottish accent variant of the Combilex lexicon \citep{richmond2009robust, richmond2010generating}.
We discard silence segments and vowels from the phone set and map the remaining phones onto one of nine classes corresponding to place of articulation:
\emph{alveolar, dental, labial, labiovelar, lateral, palatal, postalveolar, rhotic}, and \emph{velar}.
From the training data, we exclude phone instances that do not have parallel audio and ultrasound.
These instances occurred when audio started recording before the ultrasound.
For the UXTD data, there were 91 segments excluded due to early start.

We use Kaldi to extract Mel-frequency cepstral coefficients (MFCCs) for the audio signal.
MFCCs are commonly used for speech recognition, with good results reported for child speech recognition \citep{shivakumar2014improving}.
Waveforms are downsampled to 16KHz and features computed every 10ms over 25ms windows.
We keep 20 cepstral coefficients and append their respective first and second derivatives for a total of 60 features.
A high number of cepstral coefficients is helpful for child speech processing \citep{li2001automatic}.
Ultrasound frames are individually reshaped to $63\times103$ using bi-linear interpolation.
A single sample consists of an anchor frame and a set of context frames.
The anchor frame is fixed at the mid-point of each phone instance and the set of context frames are extracted over a fixed sized window of 100 ms to the left and right of the anchor frame.
Because of the different frame rates, the number of frames in the context window is different for the ultrasound and audio streams.
For the audio, each context window corresponds to 10 frames.
For ultrasound, the context window corresponds to 12 frames for Ultrasuite data and to 8 frames for TaL data.
Over each context window, we extract 5 MFCC frames and 4 ultrasound frames, with the step size set separately to account for the respective frame rates.
Figure \ref{fig:diagram-sample-build} illustrates the sample build process.

\subsection{Model architecture and training}

\begin{figure}
\includegraphics[width=\columnwidth]{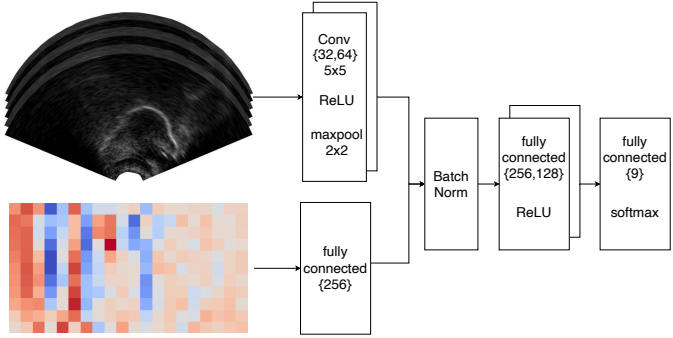}
\caption{\label{fig:diagram-model-architecture}
Convolutional neural network architecture for classifier using ultrasound and audio (MFCCs) inputs.}
\end{figure}

The adopted model architecture, illustrated in Figure \ref{fig:diagram-model-architecture}, largely follows that of earlier work \citep{ribeiro2019speaker, ribeiro2019ultrasound}.
The ultrasound stream is processed by two convolutional layers.
These layers use $5\times5$ kernels with 32 and 64 filters, respectively, and ReLU activation functions.
Each convolutional layer is followed by max-pooling with a $2\times2$ kernel.
The sequence of frames for the audio stream is flattened and processed by a fully-connected layer with rectified linear units.
When using the ultrasound and audio streams, the features are concatenated at this stage.
The batch normalized features are then processed by two fully-connected layers with ReLU activation functions and an output fully-connected layer followed by the softmax function.

Models are optimized via Stochastic Gradient Descent with minibatches of 128 samples and an L2 regularizer with weight 0.1.
We train models on the UXTD data or on the pooled TaL and UXTD data.
When using the UXTD training data, systems are optimized for 200 epochs with a learning rate of 0.1.
With the pooled dataset, systems are optimized for 50 epochs with an identical learning rate of 0.1.
After each epoch, the model is evaluated on the validation data and we keep the best model across all epochs.
We fine-tune systems trained on the pooled data on the UXTD data.
Models that are fine-tuned reduce the learning rate to 0.001 and are optimized for 100 epochs.

\subsection{Scoring}

The output of the classifier is a probability distribution over the nine places of articulation.
To score an input phone instance $x$, we consider an expected class $y$ and a competing class $c$.
The model score $s_m$ is then computed as
\begin{equation}
    s_m = \log(p(y|x)) - \log(p(c|x))
\label{eq:model_score}
\end{equation}
The expected class may be the canonical phone class, such as a velar or a rhotic.
The competing class is a possible substitution, such as an alveolar or a labiovelar approximant.
If no competing class is given, we can estimate it and compute the phone score with
\begin{equation}
c = \underset{q \in \mathcal{Q} - \{ y \}}{\arg\max}\, p(q|x)
\label{eq:model_max_score}
\end{equation}
where $\mathcal{Q}$ is the set of places of articulation considered by the model.
This method is related to the Goodness of Pronunciation score \citep{witt2000phone, hu2015improved}.
As with the combined expert score $s_c$ (Section \ref{subsec_expert_method}), 
the magnitude of the model score encodes certainty, whereas the sign encodes preference.
A positive score indicates preference for the expected class, while a negative score indicates preference for the competing class.
We simplify model and combined expert scores onto a binary correct/incorrect label $b(s)$ for error detection according to:
\begin{equation}
    b(s) =
    \begin{cases}
        0 & \text{if $s>k$} \\
        1 & \text{if $s\le k$} \\
    \end{cases}
\label{eq:model_binary}
\end{equation}
where $s$ is either $s_c$ or $s_m$ and $k$ is a configurable threshold.
Unless otherwise stated, results presented in this work use $k=0$, which treats uncertainty in the model score ($s_m=0$) as an error.
For the purposes of this analysis, uncertainty is not applicable to the combined expert score because we retain only cases that are correct or clear substitutions.

\begin{table*}[t]
\centering
\resizebox{0.89\textwidth}{!}{%
\begin{tabular}{@{}cccccccccc|cc@{}}
\textbf{Training Data} & \textbf{Alveolar} & \textbf{Dental} & \textbf{Labial} & \textbf{Labiovelar} & \textbf{Lateral} & \textbf{Palatal} & \textbf{Postalveolar} & \textbf{Rhotic} & \textbf{Velar} & \textbf{Global} &  \\ \midrule
                       & \multicolumn{10}{c}{\textit{Audio}}    \\ \midrule[.02em]

UXTD                   &  72.36\% &  46.77\%  &  64.00\%  &  52.5\%   &  84.85\%  &  64.52\%  &   68.66\% &   85.44\% &  75.62\%  &  70.81\%  \\
Joint                  &  70.56\%  &  35.14\% &  67.21\%  &  52.11\%  &  74.02\%  &  50.00\%  &   71.64\% &   88.64\% &  70.25\%  &  65.93\%   \\
Joint (+fine-tuning)   &  75.49\%  &  39.73\% &  66.67\%  &  59.42\%  &  80.56\%  &   66.67\% &   70.59\% &   85.71\%  &   77.65\% &  72.03\%   \\ \midrule[.02em]
                       & \multicolumn{10}{c}{\textit{Ultrasound}} \\ \midrule[.02em]
UXTD                   &  78.32\%  &  53.62\%   &  59.21\%   &  74.42\%   &  78.95\%   &  25.93\%  &   50.67\%  &  67.57\%  &  88.27\%   &  69.48\%   \\
Joint                  &  79.38\%  &  60.34\%   &  47.11\%   &  67.35\%   &  84.38\%   &  22.22\%  &   41.00\%  &  73.13\%  &  92.52\%   &  68.37\%   \\
Joint (+fine-tuning)   &  84.05\%  &  61.11\%   &  60.82\%   &  76.00\%   &  84.91\%   &  38.89\%  &   48.81\%  &  79.26\%  &  94.89\%   &  76.14\%   \\ \midrule[.02em]
                       & \multicolumn{10}{c}{\textit{Audio+Ultrasound}}   \\ \midrule[.02em]
UXTD                   &  80.10\%  &  81.08\%   &  83.87\%   &  82.61\%   &  90.18\%   &  65.52\%   &   58.33\%  &  74.83\%  &  89.58\%  &  80.47\%   \\
Joint                  &  83.25\%  &  76.74\%   &  83.33\%   &  63.08\%   &  73.19\%   &  66.67\%   &   76.19\%  &  91.59\%  &  93.45\%  &  81.80\%   \\
Joint (+fine-tuning)   &  87.94\%  &  76.47\%   &  87.78\%   &  72.31\%   &  91.82\%   &  58.82\%   &   75.38\%  &  94.34\%  &  95.58\%  &  \textbf{86.90\%}   \\ \midrule[.02em]
Number of samples      &  196  &  37   &  62   &  46   &  112   &  29   &   84  &  143   &  192  &   901   \\
\bottomrule
\end{tabular}%
}
\caption{Accuracy for the typically developing evaluation set across all places of articulation. Global accuracy is an average of all places of articulation, weighted by the number of samples for each class. The Joint training data denotes the pooled UXTD and TaL data. Highlighted results in bold indicate best overall performance.}
\label{tab:model-results-uxtd}
\end{table*}

\subsection{Results}

We evaluate model performance on the \textbf{typically developing} set.
Table \ref{tab:model-results-uxtd} shows accuracy results, which are computed across examples of all output classes.
Systems trained on the joint UXTD and TaL data underperform when compared with systems trained only on the UXTD data, 
even though there is more training data available.
However, fine-tuning the pre-trained joint model on the UXTD data leads to the best performance.

\begin{table}[t]
\centering
\resizebox{0.90\columnwidth}{!}{%
\begin{tabular}{@{}ccccc@{}}
\toprule
\textbf{Training data} & \textbf{Precision} & \textbf{Recall} & \textbf{F1-Score} & \textbf{Accuracy} \\ \hline
\multicolumn{5}{c}{\textit{Audio}}   \\ \midrule
UXTD                   &  0.384   &  0.524 &  0.443 &  64.1\%   \\
Joint                  &  0.417   &  0.585 &  0.487 &  66.5\%   \\ 
Joint (+fine-tuning)   &  0.393   &  0.537 &  0.454 &  64.9\%   \\ \midrule

\multicolumn{5}{c}{\textit{Ultrasound}}   \\ \midrule
UXTD                   &  0.670   &  0.842  &  0.746 &  84.4\%   \\
Joint                  &  0.702   &  0.805  &  0.750 &  85.4\%   \\ 
Joint (+fine-tuning)   &  0.677   &  0.768  &  0.720 &  83.7\%   \\ \midrule

\multicolumn{5}{c}{\textit{Audio+Ultrasound}}   \\ \midrule
UXTD                   &  0.732   &  0.866   &  \textbf{0.793} &  \textbf{87.7\%}   \\
Joint                  &  0.704   &  0.695   &  0.699 &  83.7\%   \\ 
Joint (+fine-tuning)   &  0.681   &  0.756   &  0.717 &  83.7\%   \\

\bottomrule
\end{tabular}%
}
\caption{Results for velar fronting error detection using the UXSSD velar samples rated by all annotators, except annotator 1.
Scores are computed using \enquote{velar} and \enquote{alveolar} as expected and competing classes respectively.}
\label{tab:model-results-velar}
\end{table}

Comparing systems using only one modality, accuracy results are better for ultrasound when using additional TaL data.
As expected, systems using both audio and ultrasound provide the best results.
Observing accuracy separately for each class, we observe that \emph{labial}, \emph{palatal}, \emph{postalveolar}, or \emph{rhotic} speech sounds have better results when using only audio compared to using only ultrasound.
The remaining speech sounds have better results with ultrasound tongue imaging.
Such differences are expected due to the individual characteristics of speech sounds.
For example, labial sounds do not rely on tongue movement, so they are not expected to benefit much from ultrasound tongue imaging alone.
On the other hand, velar and alveolar sounds have well-defined tongue shapes on the mid-saggital plane, so we would expect ultrasound data to be the primary contributor when identifying them.
We also observe that accuracy improves across all classes when using both modalities as input.
These results meet our expectations that ultrasound and audio complement each other well and that additional out-of-domain training data is beneficial.
Similar findings were reported on related tasks, such as speaker diarisation and word alignment of speech therapy sessions \citep{ribeiro2019ultrasound}.

Table \ref{tab:model-results-velar} shows results for \textbf{velar fronting error detection}.
These are computed over samples identified by annotators as correct velar productions or alveolar substitutions (see Figure \ref{fig:annotation-primary-secondary-matrix}).
We exclude samples rated by annotator 1, due to less than perfect agreement with other annotators.
Results are computed on $b(s)$ using the combined expert score $s_c$ and the model score $s_m$ with an expected \emph{velar} class and a competing \emph{alveolar} class.

We observe that ultrasound is more suited than audio to discriminate between velar and alveolar productions, although systems using both data streams have the best results.
There are no performance improvements to the systems using the joint dataset and fine-tuning when compared to the system using only typically developing child data.
This is an interesting observation, as results on the typically developing dataset indicate that using additional training data and fine-tuning is beneficial.
On the typically developing data, the individual accuracy for the velar and alveolar classes increases with more data and training.
Considering the system using both audio and ultrasound and comparing the UXTD and fine-tuned systems,
velar accuracy increases from 89.58\% to 95.58\% and alveolar accuracy increases from 80.10\% to 87.94\%.
The discrepancy observed between the typically developing set and velar fronting error detection could be attributed to challenges associated with speaker's data.
Speaker performance can vary substantially, particularly when using ultrasound data \citep{ribeiro2019speaker}.
This observation can be further supported by measuring agreement between model and expert binary scores for each of the eight speakers.
Using Cohen's $\kappa$ \citep{cohen1960coefficient}, models and expert scores have near perfect agreement for speakers 1 and 3 ($\kappa > 0.8$), substantial agreement for speakers 4 and 7 ($0.6 < \kappa \leq 0.8$), moderate agreement for speakers 5 and 6 ($0.4 < \kappa \leq 0.6$), and no or slight agreement ($\kappa \leq 0.2$) for speakers 2 and 8.

In Section \ref{subsec:annotator-agreement}, we reported intra- and inter-annotator agreement for the scoring of rhotic productions.
Unlike velars, expert scores for rhotics were shown to have very low inter-annotator agreement.
For this reason, results for \textbf{gliding error detection} should be interpreted carefully.
However, intra-annotator agreement was reliable and consistent for all annotators except annotator 8 (Figure \ref{fig:annotation-cohen-kappa}).
Therefore, we opt to analyse the results for gliding error detection separately for each speaker.

We note from Table \ref{tab:model-results-uxtd} that classification accuracy for rhotics and labiovelars is good across all classifiers on the typically developing evaluation set.
We observe that classification of rhotic instances benefits more from audio (85.71\%) than ultrasound (79.26\%).
The labiovelar class, on the other hand, achieves higher accuracy when using only ultrasound (76.0\%) than when using only audio (59.42\%).
Jointly using audio and ultrasound improves accuracy for rhotics (94.34\%) but not for labiovelars (72.31\%).
The average accuracy for rhotic and labiovelars when using ultrasound and audio is 78.72\% when training only on the UXTD data.
This accuracy slightly decreases when jointly training on the TaL corpus (77.34\%), but improves when fine-tuning on the UXTD data (83.33\%).
These results, however, relate to the typically developing evaluation set.
As observed with the velar case, they may not transfer in the same way to error detection.
Nevertheless, we select the fine-tuned system using both ultrasound and audio to analyse speaker-wise results for gliding error detection.

\begin{table}[t]
\centering
\resizebox{0.95\columnwidth}{!}{%
\begin{tabular}{@{}ccccccc@{}}
\toprule
\textbf{Speaker} & \textbf{N}  & \textbf{Precision} & \textbf{Recall} & \textbf{F1-Score}    & \textbf{Accuracy} & \textbf{Cohen's $\kappa$} \\ \midrule
1       & 41 & 0.778     & 0.467  & 0.583 & 75.6\%  & 0.426 \\
2       & 29 & 1.000     & 0.071  & 0.133 & 55.2\%  & 0.074 \\
3       & 36 & 0.900     & 0.474  & 0.621 & 69.4\%  & 0.404 \\
4       & 11 & 0.833     & 1.000  & 0.909 & 90.9\%  & 0.820 \\
5       & 42 & 0.964     & 0.675  & 0.794 & 66.7\%  & 0.045 \\
6       & 36 & 1.000     & 0.639  & 0.780 & 63.9\%  & 0.000 \\
7       & 45 & 0.500     & 1.000  & 0.667 & 97.8\%  & 0.656 \\
8       & 35 & 0.692     & 0.783  & 0.735 & 62.9\%  & 0.123 \\ \bottomrule
\end{tabular}%
}
\caption{Speaker-wise results for gliding error detection using the UXSSD rhotic samples identified as correct productions or gliding substitutions. Cohen's $\kappa$ is calculated on expert and model binary scores.
Results are generated by the fine-tuned system using both audio and ultrasound, with
scores computed using \enquote{rhotic} and \enquote{labiovelar} as expected and competing classes, respectively.
}
\label{tab:model-results-rhotic}
\end{table}

\begin{figure}[t]
\centering
\includegraphics[width=0.80\columnwidth]{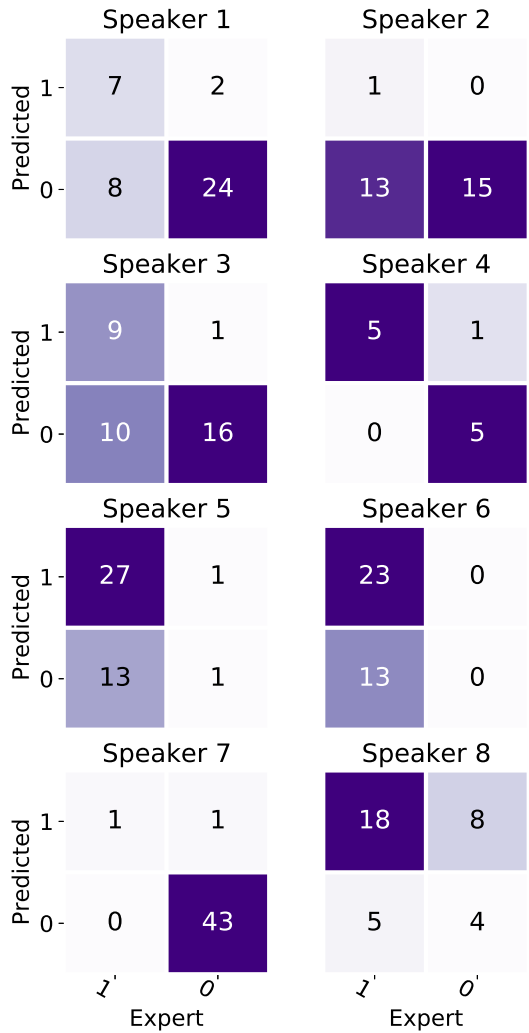}
\caption{\label{fig:rhotic_confusion_matrices}
Confusion matrices for gliding error detection produced by the fine-tuned system using audio and ultrasound for all 8 speakers in the UXSSD evaluation set. Correct rhotic productions are denoted by $0$ and gliding substitutions are denoted by $1$. }
\end{figure}

Table \ref{tab:model-results-rhotic} shows speaker-wise results for gliding error detection and Figure \ref{fig:rhotic_confusion_matrices} shows their respective confusion matrices.
We observe that the scores given by the expert annotators can vary per speaker.
For example, most samples produced by speakers 5 and 6 were marked as gliding cases by their respective annotators.
The limited number of correct productions influences the calculation of Cohen's $\kappa$, leading to poor agreement even though accuracy and F$_1$ are high.
On the other hand, most samples by speaker 7 were marked as correct instances.
The classifier appears to behave similarly with data from speakers 2 and 7, with most samples classified as correct rhotic instances.
This behaviour is in agreement with the expert for speaker 7, but not for speaker 2.
Most errors produced by the model are Type II errors (false negatives).
This might be due to the lower performance of the competing labiovelar class, as observed on the typically developing set.
Because the classifier is more confident when scoring rhotics, there is a limited number of Type I errors (false positives).
Due to the low inter-annotator agreement, it is not clear whether these differences are due to the scores provided by the annotators or due to challenges associated with speaker or recording variability.

\section{Discussion}
\label{sec:discussion}

\begin{figure*}[t!]
    \centering
    \includegraphics[width=1.00\textwidth]{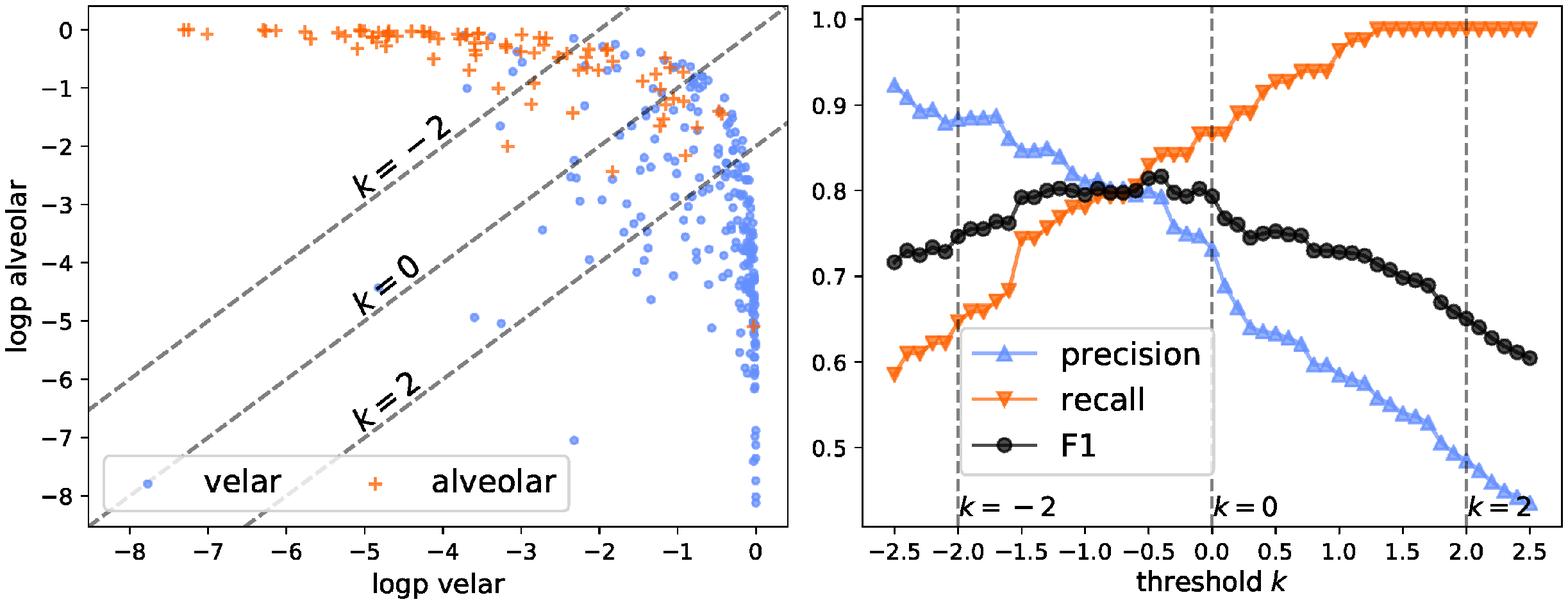}
    \caption{
Velar fronting error detection with Audio and Ultrasound system trained on UXTD data.
The figure on the left shows model scores for velar and alveolar classes with diagonal lines indicating possible thresholds.
Each sample is coloured according to its true label, as given by expert annotators, using $k=0$.
On the right, precision, recall, and F$_1$ score as a function of the threshold $k$.}
    \label{fig:results_scatter_threshold_joint}
\end{figure*}

Considering the scores provided by the expert Speech and Language Therapists, we observed that results for velar samples are consistent and reliable, meeting our initial expectations.
The high inter-annotator agreement, excluding rater 1, for the combined score ($\alpha = 0.922$) suggests that the data provided by the experienced SLTs can be used for the evaluation of automatic methods.
The scores provided on the rhotic samples, however, were less consistent and did not meet our original expectations.
There are various reasons why this might have occurred.
Because children from the Ultrasuite repository were not treated for the production of rhotics, correct and incorrect samples were  unbalanced in the annotation list.
This may have affected the judgements made by the annotators, who expected to encounter some correct productions of /r/.
Including samples from typically developing children could have mitigated this issue and helped anchor the scale for correct productions.
Additionally, /r/ is a less common target for intervention in the UK \citep{wren2016prevalence}.
This might have affected the behaviour of the experienced SLTs, who, although able to discriminate between correct and incorrect productions, have less experience when evaluating this particular speech sound clinically.
Related to this observation, there is a wide range of socially acceptable productions of /r/ in the United Kingdom \citep{scobbie2006r, lawson2011social}, which motivated the choice of gliding for analysis in this work.
Unlike velars, which have a relatively consistent tongue shape between speakers, rhotics can be produced with a wide variety of tongue shapes \citep{boyce2015articulatory}, potentially making it more difficult to judge acceptability using ultrasound.
To account for the wide range of acceptable productions of /r/, annotators could browse through a set of samples drawn from typically developing children.

With respect to automatic scoring of speech articulation errors, our results indicate that expert behaviour can be simulated with an acceptable level of accuracy for velar fronting error detection.
The best performing system correctly detected 86.6\% of all errors identified by SLTs.
Out of all the segments identified as errors, 73.2\% of those are correct.
When evaluating systems, we assumed a threshold $k=0$ to compute the final score according to Equation \ref{eq:model_binary}.
The threshold $k$ is a parameter configured by the user, allowing some control over precision and recall.
Figure \ref{fig:results_scatter_threshold_joint} illustrates model scores and the impact of $k$ on precision and recall.
As expected, most of the uncertainty with respect to the true label occurs around $k=0$.
For the system in Figure \ref{fig:results_scatter_threshold_joint}, the $F_1$ score improves from $0.793$ to $0.817$ when $k=-0.4$.

Even though changing $k$ might result in slight improvements, the ranking of the systems across all conditions remains the same.
Ultrasound tongue imaging has a positive contribution to the overall accuracy of the models, when used by itself or together with audio features.
Considering out-of-domain data, results show that model performance can be improved when pre-trained on adult speech data.
Overall performance increases further when fine-tuning models on in-domain data.
This is observed when evaluating on typically developing speech, but not when detecting errors on disordered speech data.
This discrepancy could be caused by differences in the two datasets.
The typically developing set (UXTD) and disordered speech set (UXSSD) were collected separately, with different purposes and conditions  \citep{eshky2018ultrasuite}.
This might lead to domain mismatches between training (UXTD) and test (UXSSD) data.
Although the model achieves better accuracy on typically developing data, it may not generalise to the different disordered speech domain.
Differences between the two datasets include speaker characteristics, ultrasound probe placement, or acoustic variability due to room conditions or hardware used for data collection.

Results for gliding error detection are harder to interpret due to low inter-annotator agreement.
We do observe reasonable accuracy for some speakers in the evaluation set.
However, further work should investigate primarily the processes used by annotators to score the samples.

Future work for automatic error detection should explore error processes beyond substitutions, such as insertions or deletions.
These cases could be detected using methods similar to those used for mispronunciation detection in language learning (e.g. \cite{witt2000phone, witt2012automatic, hu2015improved}).
There are various child speech corpora which could complement this type of analyses, mostly through the addition of out-of-domain acoustic data.
A recent study has identified probe placement variability in the Ultrasuite data \citep{csapo2020quantification}.
Variable probe placement could be limiting the performance of an error detection classifier.
Future work could leverage the techniques proposed by \cite{csapo2020quantification} to account for such variability at training and test time.
This could also be used to provide real-time probe placement feedback to clinicians and minimise misalignment errors.
Alternatively, to reduce domain mismatch between training and test data, unsupervised domain adversarial training \citep{ganin2016domain} could be helpful, as well as the application of in-domain data augmentation techniques \cite{shorten2019survey}.
A different direction for future work could could leverage the the temporal dependency of therapy sessions.
In this longitudinal online learning scenario \citep{karanasou2015speaker}, the SLT provides feedback in early sessions (e.g. baseline assessment session) by verifying scores given by the model.
Those verified labels are then used to improve model scores on subsequent sessions (e.g. mid-therapy or post-therapy).
This scenario could help account for annotator preferences, as well as variability due to speaker characteristics or hardware configuration.

\section{Conclusion}
\label{sec:conclusion}

We investigated the use of ultrasound tongue imaging for the detection of velar fronting and gliding of /r/ in Scottish English child speakers.
For this task, results indicate that experienced speech and language therapists have near perfect agreement when annotating the correctness of velar speech sounds, but agreement on the correctness of rhotic speech sounds is low.

For automatic error detection, out-of-domain adult data improves results on typically developing speech, but it is less useful when evaluating on disordered speech.
Results indicate that velar fronting error detection benefits more from ultrasound than audio, but we observe the best performance when using both modalities.
In terms of gliding error detection, results are harder to interpret due to low inter-annotator agreement.

Future research should explore techniques to account for speaker, session, and equipment variability with ultrasound equipment, as well as annotation preferences by speech and language therapists.
Nonetheless, the overall performance of the classifier is promising, particularly for velar fronting error detection, with good agreement with experienced speech and language therapists.
This evidence suggests there is potential for systems to be integrated into ultrasound intervention software for automatically quantifying progress during speech therapy.

\section{License and Distribution}

The pronunciation scores obtained in Section \ref{sec:expert_detection} are publicly available in the Ultrasuite Repository\footnote{\url{https://www.ultrax-speech.org/ultrasuite}} and are distributed under Attribution-NonCommercial 4.0 Generic (CC BY-NC 4.0).
A demo of the data preparation and scoring processes with the best performing model in Table \ref{tab:model-results-uxtd} is released as part of the UltraSuite code repository\footnote{\url{https://github.com/UltraSuite}} under Apache License v.2.

\section*{Acknowledgments}
We are grateful to the Speech and Language Therapists who kindly agreed to participate in our data collection.
This work was supported by the Carnegie Trust for the Universities of Scotland (Research Incentive Grant number 008585)
and the EPSRC Healthcare Partnerships grant number EP/P02338X/1 (Ultrax2020 –- \ultraxurl).

\bibliography{references}

\end{document}